\documentclass[aps,eqsecnum,preprint,floats,epsf,epsfig,nofootinbib,letter]{revtex4}
\usepackage{epsfig}
\usepackage{amsmath}

\begin{document}
\def\be{\begin{eqnarray}}
\def\en{\end{eqnarray}}
\def\non{\nonumber}
\def\ov{\overline}
\def\la{\langle}
\def\ra{\rangle}
\def\B{{\cal B}}
\def\pr{{\sl Phys. Rev.}~}
\def\prl{{\sl Phys. Rev. Lett.}~}
\def\pl{{\sl Phys. Lett.}~}
\def\np{{\sl Nucl. Phys.}~}
\def\zp{{\sl Z. Phys.}~}
\def\up{\uparrow}
\def\dw{\downarrow}
\def\lsim{ {\ \lower-1.2pt\vbox{\hbox{\rlap{$<$}\lower5pt\vbox{\hbox{$\sim$}
}}}\ } }
\def\gsim{ {\ \lower-1.2pt\vbox{\hbox{\rlap{$>$}\lower5pt\vbox{\hbox{$\sim$}
}}}\ } }

\font\el=cmbx10 scaled \magstep2{\obeylines\hfill February, 2016}

\vskip 1.5 cm

\centerline{\large\bf Heavy-Flavor-Conserving Hadronic Weak Decays }
\centerline{\large\bf of Heavy Baryons}

\bigskip
\bigskip
\centerline{\bf Hai-Yang Cheng$^1$, Chi-Yee Cheung$^1$, Guey-Lin Lin$^2$,}
\centerline{\bf Yeu-Chung Lin$^3$, Tung-Mow Yan$^4$, Hoi-Lai Yu$^1$}
\medskip
\centerline{$^1$Institute of Physics, Academia Sinica}
\centerline{Taipei, Taiwan 115, Republic of China}
\medskip
\medskip
\centerline{$^2$ Institute of Physics, National Chiao Tung University} \centerline{Hsinchu 30010, Taiwan, Republic of China}
\medskip
\medskip
\centerline{$^3$ Department of Physics, National Taiwan University}
\centerline{Taipei 10617, Taiwan, Republic of China}
\medskip
\medskip
\centerline{$^4$
Laboratory for Elementary Particle Physics, Cornell University}
\centerline{Ithaca, NY 14853, USA}

\bigskip
\bigskip
\centerline{\bf Abstract}
\bigskip
\small

More than two decades ago, we studied heavy-flavor-conserving weak decays of heavy baryons within the framework that incorporates both heavy-quark and chiral symmetries. In view of the first observation of $\Xi_b^-\to\Lambda_b^0\pi^-$ by LHCb recently, we have reexamined these decays and presented updated predictions. The predicted rates for $\Xi_b^-\to\Lambda_b^0\pi^-$ in the MIT bag and diquark models are consistent with experiment.  The major theoretical uncertainty stems from the evaluation of baryon matrix elements. The branching fraction of $\Xi_c\to\Lambda_c\pi$ is predicted to be of order $10^{-4}$. It is suppressed relative to $\B(\Xi_b\to\Lambda_b\pi)$ owing to the shorter lifetime of $\Xi_c$ relative to $\Xi_b$ and the destructive nonspectator $W$-exchange contribution. The kinematically accessible weak decays of the sextet heavy baryon $\Omega_Q$ are $\Omega_Q\to\Xi_Q\pi$.
Due to the absence of the $\B_6-\B_{\bar 3}$ transition in the heavy quark limit and the $\B_6-\B_6$  transition in the model calculations, $\Omega_Q\to\Xi_Q\pi$ vanish in the heavy quark limit.

\pagebreak

\section{Introduction}
Recently LHCb has measured for the first time the heavy-flavor-conserving and strangeness-changing weak decay $\Xi_b^-\to\Lambda_b^0\pi^-$ \cite{Aaij:2015yoy}. The relative rate is measured to be
\be
{f_{\Xi_b^-}\over f_{\Lambda_b^0}} \B(\Xi_b^-\to\Lambda_b^0\pi^-)=(5.7\pm1.8^{+0.8}_{-0.9})\times 10^{-4},
\en
where $f_{\Xi_b^-}$ and $f_{\Lambda_b^0}$ are $b\to \Xi_b^-$ and $b\to\Lambda_b^0$ fragmentation fractions, respectively. Assuming $f_{\Xi_b^-}/ f_{\Lambda_b^0}$ in the range between 0.1 and 0.3, based on the measured production rates of other strange particles relative to their non-strange counterparts \cite{Aaij:2015yoy}, the branching fraction
$\B(\Xi_b^-\to\Lambda_b^0\pi^-)$ will lie in the range from $(0.57\pm0.21)\%$ to $(0.19\pm0.07)\%$.

More than two decades ago, we studied heavy-flavor-conserving weak decays of heavy baryons within the framework that incorporates both heavy-quark and chiral symmetries \cite{ChengHFC}. Our motivation was as follows. Unlike hadronic weak decays of heavy mesons, a rigorous and reliable approach for describing the nonleptonic decays of heavy baryons does not exist. Nevertheless, there is a special class of weak decays of heavy baryons that
can be studied in a more trustworthy way, namely, heavy-flavor-conserving
nonleptonic decays. Examples are the singly
Cabibbo-suppressed decays $\Xi_Q\to\Lambda_Q\pi$ and $\Omega_Q\to\Xi_Q\pi$.
\footnote{The decays $\Omega_Q\to\Xi'_Q\pi$  are kinematically prohibited.}
The idea is simple: In these decays, only the light quarks inside the heavy baryon will participate in weak
interactions; that is, while the two light quarks undergo weak
transitions, the heavy quark behaves as a ``spectator". (An additional nonspectator contribution to the charmed baryon decay will be discussed shortly below.)
As the emitted light mesons are soft, the $\Delta S=1$ weak interactions
among light quarks can be handled by the well known short-distance
effective Hamiltonian. This special class
of weak decays can usually be tackled more reliably than the
conventional heavy baryon weak decays.
The synthesis of the heavy quark and chiral
symmetries \cite{Yan,Wise} provides a natural setting for investigating these
reactions \cite{ChengHFC}. The weak decays $\Xi_Q\to\Lambda_Q\pi$
with $Q=c,b$  were also studied in \cite{Sinha,Voloshin00,Voloshin14,Faller}.

The combined symmetries of heavy and light quarks severely
restrict the weak interactions allowed. In the symmetry limit, it
was found in \cite{ChengHFC} that ${\cal B}_{\bar 3}-{\cal B}_6$ and
${\cal B}^*_6-{\cal B}_6$ nonleptonic weak transitions cannot occur, where ${\cal B}_{\bar 3}$ and ${\cal B}_6$ are antitriplet and sextet heavy baryons, respectively, and ${\cal B}^*_6$  the spin-3/2 heavy baryon field.
Symmetries alone permit three types of transitions:  ${\cal B}_{\bar 3}-{\cal
B}_{\bar 3}$,  ${\cal B}_{6}-{\cal B}_6$ and ${\cal B}^*_6-{\cal
B}^*_6$ transitions. However, in both the MIT bag and diquark
models, only ${\cal B}_{\bar 3}-{\cal B}_{\bar 3}$ transitions
have nonzero amplitudes.

For heavy-flavor-conserving decays of charmed baryons such as $\Xi_c^0\to\Lambda_c^+\pi^-$ and $\Xi_c^+\to\Lambda_c^+\pi^0$, there is an additional contribution  arising from the $W$-exchange diagram $cs\to dc$ which is absent in the bottom-baryon sector. Hence, $\Xi_c\to\Lambda_c\pi$ can proceed through $W$-exchange which is a nonspectator effect. It is known that $W$-exchange plays a dramatic role in charmed baryon decays as it is not subject to helicity and color suppression \cite{Cheng:charmbaryon}.

In our original paper \cite{ChengHFC}, we did not provide numerical predictions for bottom baryons such as $\Xi_b\to \Lambda_b\pi$ since $\Lambda_b$ was the only bottom baryon that had been studied then.  In this work we would like to revisit heavy-flavor-conserving decays of both charmed and bottom baryons and present updated predictions.

This work is organized as follows. In Sec. II we set up the formalism suitable for analyzing the heavy-flavor-conserving decays of heavy baryons. Some model calculations are presented in Sec. III.
Sec. IV gives our conclusion and discussions. Appendix A is devoted to the MIT bag model evaluation of baryon matrix elements.

\section{Formalism}
The effective $\Delta S=1$ weak Hamiltonian at the scale $\mu=m_c$ reads \cite{Buchalla}
\footnote{The full effective $\Delta S=1$ Hamiltonian at $\mu<m_c$ is given by \cite{Buchalla}
\be
H_{\rm eff}^{\Delta S=1}={G_F\over \sqrt{2}}V^*_{ud}V_{us}\sum_{i=1}^{10}(z_i+\tau y_i)O_i+h.c., \non
\en
with $\tau=-(V^*_{td}V_{ts})/(V^*_{ud}V_{us})$. The numerical results of $z_i$ and $y_i$ can be found in \cite{Buchalla}. At the scale $\mu=m_c$, $y_1=y_2=0$ and $z_3,\cdots,z_{10}$ are numerically irrelevant relative to $z_1$ and $z_2$. Hence, we change the notation from $z_1,z_2$ to $c_1,c_2$.}
\be  \label{eq:H}
H_{\rm eff}^{\Delta S=1}={G_F\over \sqrt{2}}V^*_{ud}V_{us}(c_1O_1+c_2O_2)+h.c.,
\en
and the four-quark operators are given by
\be
O_1=(\bar du)(\bar us), \qquad O_2=(\bar ds)(\bar uu),
\en
with $(\bar q_1q_2)\equiv \bar q_1\gamma_\mu(1-\gamma_5)q_2$.
We shall use $c_1=1.216$ and $c_2=-0.415$ obtained at the scale $\mu=m_c$ and in the naive dimensional regularization scheme with $\Lambda^{(4)}_{\ov {\rm MS}}=325$ MeV \cite{Buchalla}.
As pointed out in \cite{Voloshin00}, there is an additional ``nonspectator" $W$-exchange contribution to heavy-flavor-conserving decays of charmed baryons governed by
\be \label{eq:Hc}
\tilde H_{\rm eff}^{\Delta S=1}={G_F\over \sqrt{2}}V^*_{cd}V_{cs}(c_1\tilde O_1+c_2\tilde O_2)+h.c.,
\en
with $\tilde O_1=(\bar dc)(\bar cs)$ and $\tilde O_2=(\bar cc)(\bar d s)$.

The general $|\Delta S|=1$ effective weak chiral Lagrangian responsible for heavy-flavor-conserving hadornic weak decays of heavy baryons is given by \cite{ChengHFC}
\be \label{eq:L}
{\cal L}^{\Delta S=1} &=& h_1{\rm tr}(\ov \B_{\bar 3}\xi^\dagger\lambda_6 \xi \B_{\bar 3})+ h_2{\rm tr}(\ov \B_{6}\xi^\dagger\lambda_6 \xi \B_{6}) \non \\
 &+& h_3{\rm tr}(\ov \B_{6}\xi^\dagger\lambda_6 \xi \B_{\bar 3})+h.c. +h_4{\rm tr}(\ov \B_6^{*\mu}\xi^\dagger\lambda_6 \xi \B_{6\mu}^*),
\en
where $\lambda_6$ is one of the SU(3) generators
\be
\lambda_6=\left(
\begin{array}{ccc}
0 & 0 & 0 \\
0 & 0 &  1 \\
0 & 1 & 0 \\
\end{array}
\right),
\en
and the pseudoscalar meson field is described by the $\xi$ term
\be
\xi={\rm exp}\left(i{M\over f_\pi}\right),
\en
with $f_\pi=132$ MeV and
\be
M &=& \left(
\begin{array}{ccc}
{\pi^0\over\sqrt{2}}+{\eta\over\sqrt{6}} & \pi^+ & K^+ \\
\pi^- & -{\pi^0\over\sqrt{2}}+{\eta\over\sqrt{6}} &  K^0 \\
K^- & \ov K^0 & -\sqrt{2\over 3}\eta \\
\end{array}
\right).
\en
The antisymmetric antitriplet $\B_{\bar 3}$ and symmetric sextet $\B_6$ in Eq. (\ref{eq:L}) are dictated by the matrices
\be
\B_{\bar 3} &=& \left(
\begin{array}{ccc}
0 & \Lambda_b^0 & \Xi_b^0 \\
-\Lambda_b^0 & 0 &  \Xi_b^- \\
-\Xi_b^0 & -\Xi_b^- & 0 \\
\end{array}
\right),  \qquad
\left(
\begin{array}{ccc}
0 & \Lambda_c^+ & \Xi_c^+ \\
-\Lambda_c^+ & 0 &  \Xi_c^0 \\
-\Xi_c^+ & -\Xi_c^0 & 0 \\
\end{array}
\right), \non \\
\B_6 &=& \left(
\begin{array}{ccc}
\Sigma_b^+ & {1\over\sqrt{2}}\Sigma_b^0 & {1\over\sqrt{2}}\Xi_b^{'0} \\
{1\over\sqrt{2}}\Sigma_b^0 & \Sigma_b^- & {1\over\sqrt{2}}\Xi_b^{'-}  \\
{1\over\sqrt{2}}\Xi_b^{'0}  & {1\over\sqrt{2}}\Xi_b^{'-} & \Omega_b^- \\
\end{array}
\right),  \qquad
\left(
\begin{array}{ccc}
\Sigma_c^{++} & {1\over\sqrt{2}}\Sigma_c^+ & {1\over\sqrt{2}}\Xi_b^{'+} \\
{1\over\sqrt{2}}\Sigma_c^+ & \Sigma_c^0 & {1\over\sqrt{2}}\Xi_c^{'0}  \\
{1\over\sqrt{2}}\Xi_c^{'+}  & {1\over\sqrt{2}}\Xi_c^{'0} & \Omega_c^0 \\
\end{array}
\right),
\en
for bottom and charmed baryons, respectively.

The combined symmetries of heavy quark and chiral symmetries severely restrict the weak transitions allowed. We have shown in \cite{ChengHFC} that $h_3=0$ and $h_2=-h_4\equiv h'$ in the symmetry limit. As a consequence,
\be \label{eq:L}
{\cal L}^{\Delta S=1} &=& h\,{\rm tr}(\ov \B_{\bar 3}\xi^\dagger\lambda_6 \xi \B_{\bar 3})+ h'{\rm tr}(\ov \B_{6}\xi^\dagger\lambda_6 \xi \B_{6})
  -h'{\rm tr}(\ov \B_6^{*\mu}\xi^\dagger\lambda_6 \xi \B_{6\mu}^*),
\en
with $h\equiv h_1$.
Hence, there cannot be $\B_{\bar 3}-\B_6$ and $\B_6^*-\B_6$ weak transitions in the heavy quark limit. Symmetries alone permit only three types of transitions: $\B_{\bar 3}-B_{\bar 3}$, $\B_6-\B_6$ and $\B_6^*-\B_6^*$ transitions. The heavy quark symmetry predicts that the couplings $h$ and $h'$ are independent of heavy quark masses. Furthermore, it was pointed out in \cite{ChengHFC} that $h'$ vanishes in the MIT-bag-model and diquark-model calculations.

The general amplitude for $\B_i\to \B_f+P$ is given by
\begin{eqnarray} \label{eq:Amp}
M(\B_i\to \B_f+P)=i\bar u_f(A-B\gamma_5)u_i,
\end{eqnarray}
where $A$ and $B$ are the $S$- and $P$-wave amplitudes, respectively. In the heavy quark limit, the diquark of the antitriplet baryon $\B_{\bar 3}$ is a scalar diquark with $J^P=0^+$, while diquark of the $\B_6$ is an axial-vector diquark with $J^P=1^+$. Therefore, the weak diquark transition is $0^+\to 0^+ + 0^-$ for  $\B_{\bar 3}\to \B_{\bar 3}+P$ and $1^+\to 1^+ + 0^-$ for $\B_6\to \B_6+P$. Based on the conservation of angular momentum, it is easily seen that the parity-conserving $P$-wave amplitude vanishes in $\B_{\bar 3}\to \B_{\bar 3}+P$ decays, but not so in $\B_6\to \B_6+P$ decays.

The general amplitude consists of factorizable and nonfactorizable ones
\be
M(\B_i\to \B_f+P)=M(\B_i\to \B_f+P)^{\rm fact}+M(\B_i\to \B_f+P)^{\rm nf}.
\en
While the factorizable amplitude vanishes in the soft meson limit, the nonfactorizable is not. We consider the latter amplitude first and take $\Xi_b^-\to\Lambda_b^0\pi^-$ as an example. The $S$-wave parity-violating amplitude can be evaluated using the chiral Lagrangian or current algebra. We find from the chiral Lagrangian Eq. (\ref{eq:L}) that
\be
M(\Xi_b^-\to\Lambda_b^0\pi^-)_{\rm S~wave}=-\la \Lambda_b^0\pi^-|{\cal L}^{\Delta S=1}|\Xi_b^-\ra=-{ih\over f_\pi}\bar u_{\Lambda_b}u_{\Xi_b},
\en
and hence the $S$-wave amplitude
\be \label{eq:S}
A(\Xi_b^-\to\Lambda_b^0\pi^-)=-{h\over f_\pi}=-{1\over f_\pi}\la \Lambda_b^0|{\cal L}^{\Delta S=1}|\Xi_b^0\ra.
\en

For the short-distance effective Hamiltonian (\ref{eq:H}), one can use current algebra to evaluate the $S$-wave amplitude. The soft pion relation leads to
\be
\la \pi^-\Lambda_b^0|H_{\rm eff}^{\rm PV}|\Xi_b^-\ra = -{i\over f_\pi}\la \Lambda_b^0|[Q_5^+, H_{\rm eff}^{\rm PV}]|\Xi_b^-\ra
= {i\over f_\pi}\la \Lambda_b^0|H_{\rm eff}^{\rm PC}|\Xi_b^0\ra.
\en
Therefore, the $S$-wave amplitude reads
\be
A(\Xi_b^-\to\Lambda_b^0\pi^-)={1\over f_\pi}\la \Lambda_b^0|H_{\rm eff}|\Xi_b^0\ra
\en
in agreement with Eq. (\ref{eq:S}) derived from the chiral Lagrangian.

The $P$-wave amplitude of the $\B_{\bar 3}\to \B_{\bar 3}+P$ decay vanishes in the heavy quark limit. The parity-conserving pole diagrams vanish due to the vanishing $\B_{\bar 3}\B_{\bar 3}\pi$ coupling and $\B_6-\B_{\bar 3}$ weak transition in heavy quark limit.
In \cite{ChengHFC} we have considered possible contributions to the $P$-wave amplitude arising from $\Xi_Q$ and $\Xi'_Q$ mixing. However, this is only one of many possible $1/m_Q$ corrections. For reason of consistency, we should work in the heavy quark limit.

Due to the absence of the $\B_6-\B_{\bar 3}$ transition in the heavy quark limit and $\B_6-\B_6$ and $\B_6^*-\B_6^*$ transitions in the model calculations, it is easily seen that there is no nonfactorizable contribution to $\Omega_Q\to\Xi_Q\pi$.

There are factorizable contributions which vanish in the soft meson limit
\be
\la \pi^-\Lambda_c^+|H_{\rm eff}|\Xi_c^0\ra^{\rm fac} &=& {G_F\over\sqrt{2}}V_{ud}^*V_{us}\left(c_1+{c_2\over 3}\right)\la \pi^-|(\bar du)|0\ra\la \Lambda_c^+|(\bar us)|\Xi_c^0\ra, \non \\
\la \pi^0\Lambda_c^+|H_{\rm eff}|\Xi_c^+\ra^{\rm fac} &=& {G_F\over\sqrt{2}}V_{ud}^*V_{us}\left(c_2+{c_1\over 3}\right)\la \pi^0|(\bar uu)|0\ra\la \Lambda_c^+|(\bar ds)|\Xi_c^+\ra.
\en
In terms of the form factors defined by
\be
\la \Lambda_c^+|(\bar u s)|\Xi_c^0\ra &=& \bar u_{\Lambda_c}[f_1^{\Lambda_c\Xi_c}\gamma_\mu+f_2^{\Lambda_c\Xi_c}i\sigma_{\mu\nu}q^\nu+f_3^{\Lambda_c\Xi_c}q_\mu \non \\
&& -g_1^{\Lambda_c\Xi_c}\gamma_\mu\gamma_5-g_2^{\Lambda_c\Xi_c}i\sigma_{\mu\nu}q^\nu\gamma_5-g_3^{\Lambda_c\Xi_c}q_\mu\gamma_5]u_{\Xi_c},
\en
we obtain
\be
\la \pi^-\Lambda_c^+|H_{\rm eff}|\Xi_c^0\ra^{\rm fac} &=& -i{G_F\over\sqrt{2}}V_{ud}^*V_{us}\left(c_1+{c_2\over 3}\right)f_\pi \bar u_{\Lambda_c}\huge[(m_{\Xi_c}-m_{\Lambda_c})f_1^{\Lambda_c\Xi_c} \non \\
&& +(m_{\Xi_c}+m_{\Lambda_c})g_1^{\Lambda_c\Xi_c}\gamma_5\huge]u_{\Xi_c}.
\en
Since the form factor $g_1^{\Lambda_c\Xi_c}$ vanishes in the heavy quark limit \cite{Yan}, only the $S$-wave amplitude receives a factorizable contribution
\be
A(\Xi_c^0\to\Lambda_c^+\pi^-)=-{G_F\over\sqrt{2}}V_{ud}^*V_{us}\left(c_1+{c_2\over 3}\right)f_\pi (m_{\Xi_c}-m_{\Lambda_c})f_1^{\Lambda_c\Xi_c}.
\en
Likewise,
\be
A(\Xi_c^+\to\Lambda_c^+\pi^0) &=& -{G_F\over 2}V_{ud}^*V_{us}\left(c_2+{c_1\over 3}\right)f_\pi (m_{\Xi_c}-m_{\Lambda_c})f_1^{\Lambda_c\Xi_c}.
\en

Heavy-flavor-conserving decays of charmed baryons receive additional non-spectator $W$-exchange contribution from Eq. (\ref{eq:Hc}):
\be
A(\Xi_c^0\to\Lambda_c^+\pi^-)&=& {1\over f_\pi}\la \Lambda_c^+|H_{\rm eff}|\Xi_c^+\ra+{1\over f_\pi}\la \Lambda_c^+|\tilde H_{\rm eff}|\Xi_c^+\ra \non \\
&-&{G_F\over\sqrt{2}}V_{ud}^*V_{us}\left(c_1+{c_2\over 3}\right)f_\pi (m_{\Xi_c}-m_{\Lambda_c})f_1^{\Lambda_c\Xi_c}, \non \\
A(\Xi_c^+\to\Lambda_c^+\pi^0)&=& {1\over \sqrt{2}f_\pi}\la \Lambda_c^+|H_{\rm eff}|\Xi_c^+\ra+{1\over \sqrt{2}f_\pi}\la \Lambda_c^+|\tilde H_{\rm eff}|\Xi_c^+\ra \non \\
&-&{G_F\over 2\sqrt{2}}V_{ud}^*V_{us}\left(c_2+{c_1\over 3}\right)f_\pi (m_{\Xi_c}-m_{\Lambda_c})f_1^{\Lambda_c\Xi_c}.
\en
To proceed, we write
\be
c_1O_1+c_2O_2={1\over 2}(c_1+c_2)(O_1+O_2)+{1\over 2}(c_1-c_2)(O_1-O_2).
\en
Since $O_1\pm O_2$ is symmetric (antisymmetric) in color indices, only $O_1-O_2$ contributes. Consequently,
\be
\la \Lambda_c^+|H_{\rm eff}|\Xi_c^+\ra &=& {G_F\over 2\sqrt{2}}V_{ud}^*V_{us}(c_1-c_2)X, \non \\
\la \Lambda_c^+|\tilde H_{\rm eff}|\Xi_c^+\ra &=& {G_F\over 2\sqrt{2}}V_{cd}^*V_{cs}(c_1-c_2)Y,
\en
with
\be \label{eq:Y2Z2}
X &\equiv& \la \Lambda_c^+|(\bar du)(\bar us)-(\bar uu)(\bar ds)|\Xi_c^+\ra , \non \\
Y &\equiv& \la \Lambda_c^+|(\bar dc)(\bar cs)-(\bar cc)(\bar ds)|\Xi_c^+\ra.
\en

To sum up, the $S$-wave amplitudes of heavy-flavor-conserving decays of heavy baryons read
\be
A(\Xi_c^0\to\Lambda_c^+\pi^-) &=& {G_F\over\sqrt{2} f_\pi}V_{ud}^*V_{us}\left[ {1\over 2}(c_1-c_2)\left(X-Y\right)-\left( c_1+{c_2\over 3}\right)f_\pi^2(m_{\Xi_c}-m_{\Lambda_c})f_1^{\Lambda_c\Xi_c}\right], \non \\
A(\Xi_c^+\to\Lambda_c^+\pi^0) &=& {G_F\over 2\sqrt{2} f_\pi}V_{ud}^*V_{us}\left[{1\over 2}(c_1-c_2)\left(X-Y\right)-\left( c_2+{c_1\over 3}\right)f_\pi^2(m_{\Xi_c}-m_{\Lambda_c})f_1^{\Lambda_c\Xi_c}\right], \non \\
A(\Xi_b^-\to\Lambda_b^0\pi^-) &=& {G_F\over\sqrt{2} f_\pi}V_{ud}^*V_{us}\left[ {1\over 2}(c_1-c_2)X-\left( c_1+{c_2\over 3}\right)f_\pi^2(m_{\Xi_b}-m_{\Lambda_b})f_1^{\Lambda_b\Xi_b}\right], \non \\
A(\Xi_b^0\to\Lambda_b^0\pi^0) &=& {G_F\over 2\sqrt{2} f_\pi}V_{ud}^*V_{us}\left[ {1\over 2}(c_1-c_2)X-\left( c_2+{c_1\over 3}\right)f_\pi^2(m_{\Xi_b}-m_{\Lambda_b})f_1^{\Lambda_b\Xi_b}\right].
\en
In terms of the amplitude given in Eq. (\ref{eq:Amp}), the decay rate of $\B_i\to\B_f+P$ is given by
\be
\Gamma &=& {p_c\over 8\pi}\left\{ {(m_i+m_f)^2-m_P^2\over m_i^2} |A|^2+{(m_i-m_f)^2-m_P^2\over m_i^2} |B|^2\right\}, \non \\
&=& {p_c\over 8\pi}\left\{ {(m_i+m_f)^2-m_P^2\over m_i^2} |A|^2+{4p_c^2\over (m_i+m_f)^2-m_P^2 } |B|^2\right\},
\en
with $p_c$ being the c.m. three-momentum in the rest frame of $\B_i$.
As noticed in passing, the $P$-wave amplitudes of interest vanish in the heavy quark limit.

\section{Model Calculations}

The four-quark matrix element $\langle \Lambda_c^+|{\cal H}_{\rm eff}^{\rm PC}|\Xi_c^+\rangle$ was evaluated in \cite{ChengHFC} using two different models: the MIT bag model \cite{MIT} and the diquark model \cite{diquark}. We first consider the bag model evaluation.  The baryon matrix elements $X$ and $Y$ defined in Eq. (\ref{eq:Y2Z2}) are obtained from Appendix A to be
\be \label{eq:XY}
X =  32\pi Y_2, \qquad  Y=8\pi Z_2,
\en
with
\be  \label{eq:Y2}
Y_2 &=& \int_0^R r^2 dr(u_du_u+v_dv_u)(u_su_u+v_sv_u), \non \\
Z_2 &=& \int_0^R r^2 dr(u_du_c+v_dv_c)(u_su_c+v_sv_c),
\en
where $u(r)$ and  $v(r)$ are the large and small components of the quark wave function, respectively, given in Eq. (\ref{eq:uv}).
In the bag model the form factors have the expression \cite{ChengHFC}
\be
f_1^{\Lambda_c\Xi_c}=f_1^{\Lambda_b\Xi_b}=-4\pi\int_0^R r^2dr(u_uu_s+v_uv_s).
\en
Numerically,
\be
Y_2=1.66\times 10^{-4}\,{\rm GeV}^3, \qquad
Z_2=2.11\times 10^{-4}\,{\rm GeV}^3, \qquad f_1^{\Lambda_c\Xi_c}=-0.985\,,
\en
where we have employed the following bag parameters
\be
m_u=m_d=0, \quad m_s=0.279~{\rm GeV}, \quad m_c=1.551~{\rm GeV}, \quad R=5~{\rm GeV}^{-1}.
\en

For the evaluation of the baryon matrix elements in the diquark model, we notice that through the Fierz identify (see Eq. (100) of \cite{Nieves:2003in})
\be
(\bar q_1q_3)(\bar q_2q_4)=2\bar q_{1i}(1+\gamma_5)q_{2j}^c\,\bar q^c_{4j}(1-\gamma_5)q_{3i}
\en
with $q^c$ being the charge conjugate quark field,
the operator $O_-=(\bar du)(\bar us)-(\bar uu)(\bar ds)$ can be recast to a local diquark-current form \cite{Stech:1987fa}
\be
O_-=2(du)^\dagger_{\bar 3}(us)_{\bar 3},
\en
where $(us)_{\bar 3}\equiv \epsilon_{ijk}\bar u^c_j(1-\gamma_5)s_k$ with $i,j,k,l$ being the color indices is a color-antitriplet scalar and pseudoscalar diquark. Note there is no color-sextet diquark-current contribution to $O_-$. In terms of the diquark currents, the expression of $O_-$ has a simple interpretation: It annihilates a scalar or pseudoscalar $(us)$ diquark in the initial  baryon and creates a scalar or pseudoscalar $(du)$ diquark in the final baryon. Following \cite{diquark} to define the diquark decay constant $g_{qq'}$
\be
\la 0|\epsilon_{ijk}\bar q^c_j\gamma_5 q'_k|(qq')^{0^+}_l\ra=\sqrt{2\over 3} \delta_{il}\,g_{qq'}
\en
for a $0^+$ scalar diquark and applying the vacuum insertion approximation, we obtain
\be
X\equiv \la \Lambda_c^+|(\bar du)(\bar us)-(\bar uu)(\bar ds)|\Xi_c^+\ra={2\over 3\, m_{\rm di}} g_{du}g_{us},
\en
where $m_{\rm di}$ is the diquark mass. We shall follow \cite{diquark} to use
\be
(c_1-c_2)g_{du}g_{us}=0.066\pm0.013~{\rm GeV}^4,
\en
which is practically scale independent. As for the diquark mass, we choose \cite{ChengHFC}
\be
m_{\rm di}=m_{\Lambda_c^+}-m_c\approx 785~{\rm MeV}
\en
with the constituent charm quark mass $\sim 1.5$ GeV.

There is an alternative approach for the evaluation of $X$. As shown in \cite{Voloshin00}, this baryon matrix element can be related to the matrix elements $x$ and $y$ defined by
\be
x=-\la \Lambda_c^+|(\bar c\gamma_\mu c)(\bar d\gamma^\mu s)|\Xi_c^+\ra, \qquad
y=-\la \Lambda_c^+|(\bar c_i\gamma_\mu c_k)(\bar d_k\gamma^\mu s_i)|\Xi_c^+\ra,
\en
which in turn can be extracted from the lifetime differences of charmed baryons. For details, see Refs. \cite{Voloshin00,Voloshin14}.

\begin{table}[t]
\caption{Masses and lifetimes of charmed and bottom baryons in units of MeV and sec, respectively, taken from \cite{PDG}.}
\label{tab:Lifetimes}
\begin{center}
\begin{tabular}{|c | c c c c|} \hline
 & $\Lambda_c^+$ & $\Xi_c^+$ & $\Xi_c^0$ & $\Omega_c^0$ \\
 \hline
Mass & $2286.46\pm0.14$ & $2467.93^{+0.28}_{-0.40}$ & $2470.85^{+0.28}_{-0.40}$ & $2695.2\pm1.7$ \\
~Lifetime~ & $(200\pm6)\times 10^{-15}$ & $(442\pm26)\times 10^{-15}$ & $(112^{+13}_{-10})\times 10^{-15}$ & $(69\pm12)\times 10^{-15}$ \\
 \hline
& $\Lambda_b^0$ & $\Xi_b^0$ & $\Xi_b^-$ & $\Omega_b^-$ \\
 \hline
Mass & $5619.51\pm0.23$ & $5794.4\pm1.2$ & $5791.8\pm0.5$ & $6048.0\pm1.9$ \\
~Lifetime~ & $(1.466\pm0.010)\times 10^{-12}$ & $(1.560\pm0.040)\times 10^{-12}$ & $(1.464\pm0.031)\times 10^{-12}$ & $(1.57^{+0.23}_{-0.20})\times 10^{-12}$ \\
 \hline
\end{tabular}
\end{center}
\end{table}

\begin{table}[t]
\caption{The magnitude of the $S$-wave amplitudes $|A|$ and branching fractions of heavy-flavor-conserving charmed and bottom baryon decays. The predictions based on the MIT bag model (diquark model) are exhibited in the first (second) entry.}
\label{tab:BR}
\begin{center}
\begin{tabular}{|l c c | l c c|} \hline
~~Mode & $|A|$ & $\B$ & ~~Mode & $|A|$ & $\B$ \\
\hline
~~$\Xi_c^0\to \Lambda_c^+\pi^-$~~ & $1.7\times 10^{-7}$ & ~~$0.87\times 10^{-4}$~~  & ~~$\Xi_b^-\to\Lambda_b^0\pi^-$~~ & $2.3\times 10^{-7}$ & ~~$2.0\times 10^{-3}$~~ \\
 & & &  & $4.3\times 10^{-7}$ & ~~$6.9\times 10^{-3}$~~ \\
~~$\Xi_c^+\to \Lambda_c^+\pi^0$~~ & $0.9\times 10^{-7}$ & ~~$0.93\times 10^{-4}$~~  & ~~$\Xi_b^0\to\Lambda_b^0\pi^0$~~ & $1.3\times 10^{-7}$ & ~~$5.9\times 10^{-4}$~~ \\
 & & & & $2.7\times 10^{-7}$ & ~~$2.5\times 10^{-3}$~~ \\
 \hline
\end{tabular}
\end{center}
\end{table}

Likewise, for the baryon matrix element $Y$ we have
\be
Y=\la\Lambda_c^+|2(dc)^\dagger_{\bar 3}(cs)_{\bar 3}|\Xi_c^+\ra.
\en
In the heavy quark limit, the charmed baryon is a bound state of the $c$ quark and the diquark $(qq')$. Hence, the diquark model is not suitable for estimating the matrix element $Y$.

With the measured lifetimes and masses of heavy baryons (see Table \ref{tab:Lifetimes}), the predicted branching fractions of heavy-flavor-conserving charmed and bottom baryon decays are summarized in Table \ref{tab:BR} calculated using the MIT bag model and the diquark model. Since the baryon matrix element $Y$ cannot be reliably estimated based on the diquark model, predictions for charmed baryons are made only for the bag model. The calculated $\B(\Xi_b^-\to \Lambda_b^0\pi^-)\approx (2.0-6.9)\times 10^{-3}$ are consistent with the LHCb measurement. The experimental branching fraction suffers from the large uncertainty with the ratio $R=f_{\Xi_b^-}/f_{\Lambda_b^0}$. As pointed out in \cite{Voloshin:2015xxa}, the uncertainty in $R$ can be greatly reduced by measuring, for example, $\Xi_b^-\to J/\psi\Xi^-$ and $\Lambda_b^0\to J/\psi \Lambda$ at LHCb. These two modes are related via SU(3) symmetry.

The branching fraction of $\Xi_c\to\Lambda_c\pi$ is predicted to be of order $10^{-4}$. Although its $S$-wave amplitude is of similar size as that of $\Xi_b\to \Lambda_b\pi$ (see Table \ref{tab:BR}), its rate is smaller owing to the shorter lifetime of $\Xi_c$ relative to $\Xi_b$. Moreover, $\Gamma(\Xi_c\to\Lambda_c\pi)$ is suppressed by the $W$-exchange contribution.

In this work we did not calculate $\Omega_Q\to\Xi_Q\pi$ rates as they vanish in the heavy quark limit. Some crude estimates given in \cite{Faller} indicate $\B(\Omega_b\to\Xi_b\pi)\sim {\cal O}(10^{-6})$ and $\B(\Omega_c\to\Xi_c\pi)< {\cal O}(10^{-6})$.

\section{Conclusion}
In light of the first observation of $\Xi_b^-\to\Lambda_b^0\pi^-$ by LHCb recently, we have reexamined the heavy-flavor-conserving decays of heavy baryons and presented updated predictions. The predicted rates for $\Xi_b^-\to\Lambda_b^0\pi^-$ in the MIT bag model and the diquark model are consistent with experiment. The major theoretical uncertainty stems from the evaluation of baryon matrix elements.
The branching fraction of $\Xi_c\to\Lambda_c\pi$ is predicted to be of order $10^{-4}$. It is suppressed relative to $\B(\Xi_b\to\Lambda_b\pi)$ owing to the shorter lifetime of $\Xi_c$ relative to $\Xi_b$ and the destructive nonspectator $W$-exchange contribution. The kinematically allowed weak decays for the sextet heavy baryon $\Omega_Q$ are $\Omega_Q\to\Xi_Q\pi$.
Due to the absence of the $\B_6-\B_{\bar 3}$ transition in the heavy quark limit and the $\B_6-\B_6$  transition in the model calculations, $\Omega_Q\to\Xi_Q\pi$ vanish in the heavy quark limit.

Finally, it is worth remarking that, analogous to the
heavy-flavor-conserving nonleptonic weak decays discussed in
this work, there is a special class of weak radiative decays in
which heavy flavor is conserved, for example, $\Xi_Q \to
\Lambda_Q \gamma$ and $\Omega_Q \to \Xi_Q \gamma$.   However, the dynamics of these radiative decays is more
complicated than that of their counterpart in nonleptonic weak decays,
e.g., $\Xi_Q \to \Lambda_Q \pi$. In any event, it merits an investigation.

\section{Acknowledgments}
This research was supported in part by the Ministry of Science and Technology of R.O.C. under Grant
Nos. 104-2112-M-001-022 and 104-2112-M-009-021.

\appendix

\section{Matrix elements in the bag model}
Consider the four-quark operator $O_-=(\bar q_1q_3)(\bar q_2q_4)$. For the evaluation of the baryon  matrix element of $O_-$ in the bag model, see \cite{Cheng:charmbaryon}. The operator can be written as
$O_-=6(\bar q_1q_3)_1(\bar q_2q_4)_2$, where the superscript $i$ indicates that the quark operator acts only on the $i$th quark in the baryon wave function.
In the bag model the parity-conserving matrix elements have the expression \cite{Cheng:charmbaryon}
\be
\int r^2dr\la q_1q_2|(\bar q_1q_3)_1(\bar q_2q_4)_2|q_3q_4\ra &=& (-X_1+X_2)-{1\over 3}(X_1+3X_2){\boldsymbol\sigma}_1\cdot {\boldsymbol\sigma}_2, \non \\
\int r^2dr\la q_1q_2|(\bar q_1q_4)_1(\bar q_2q_3)_2|q_3q_4\ra &=& (X_1+X_2)-{1\over 3}(-X_1+3X_2){\boldsymbol\sigma}_1\cdot{\boldsymbol\sigma}_2,
\en
with
\be
X_1 &=& \int_0^R r^2dr(u_1v_2-v_1u_2)(u_3v_4-v_3u_4), \non \\
X_2 &=& \int_0^R r^2dr(u_1u_2+v_1v_2)(u_3u_4+v_3v_4),
\en
where $R$ is the radius of the bag and $u(r), v(r)$ are the large and small components of the quark wave function, respectively, defined by
\be \label{eq:uv}
\psi=
\left(
\begin{array}{c}
iu(r)\chi \\
v(r){\boldsymbol\sigma}\cdot{\bf \hat{r}}\chi \\
\end{array}
\right).
\en

Applying the relation
\be
{\boldsymbol\sigma}_1\cdot{\boldsymbol\sigma}_2={1\over 2}(\sigma_{1+}\sigma_{2-}+\sigma_{1-}\sigma_{2+})+\sigma_{1z}\sigma_{2z}
\en
and the wave functions
\be
\Lambda_c^+ &=& -{1\over \sqrt{12}}\left[u^\uparrow d^\downarrow c^\uparrow-u^\dw d^\up c^\up-d^\up u^\dw c^\up+d^\dw u^\up c^\up+(13)+(23)\right], \non \\
\Xi_c^+ &=& {1\over \sqrt{12}}\left[u^\uparrow s^\downarrow c^\uparrow-u^\dw s^\up c^\up-s^\up u^\dw c^\up+s^\dw u^\up c^\up+(13)+(23)\right],
\en
with obvious notation for permutation of quarks,
it is straightforward to show that
\be \label{eq:Y2Y1}
\la \Lambda_c^+|(\bar du)(\bar us)|\Xi_c^+\ra
&=& 6(4\pi)\la \Lambda_c^+|b^\dagger_{1d}b_{1u}b^\dagger_{2u}b_{2s}[Y_2(1- {\boldsymbol\sigma}_1\cdot {\boldsymbol\sigma}_2)-Y_1(1+{1\over 3}{\boldsymbol\sigma}_1\cdot {\boldsymbol\sigma}_2)|\Xi_c^+\ra \non \\
&=& 16\pi Y_2, \non \\
\la \Lambda_c^+|(\bar uu)(\bar ds)|\Xi_c^+\ra
&=& 6(4\pi)\la \Lambda_c^+|b^\dagger_{1u}b_{1u}b^\dagger_{2d}b_{2s}[Y_2(1- {\boldsymbol\sigma}_1\cdot {\boldsymbol\sigma}_2)+Y_1(1+{1\over 3}{\boldsymbol\sigma}_1\cdot {\boldsymbol\sigma}_2)|\Xi_c^+\ra \non \\
&=& -16\pi Y_2,
\en
with $Y_2$ given by Eq. (\ref{eq:Y2}). Likewise,
\be \label{eq:Z2Z1}
\la \Lambda_c^+|(\bar dc)(\bar cs)|\Xi_c^+\ra=4\pi(Z_2-Z_1), \qquad
\la \Lambda_c^+|(\bar cc)(\bar ds)|\Xi_c^+\ra=-4\pi(Z_2+Z_1),
\en
with
\be
Z_1 &=& \int_0^R r^2 dr(u_dv_c-v_du_c)(u_sv_c-v_su_c), \non \\
Z_2 &=& \int_0^R r^2 dr(u_du_c+v_dv_c)(u_su_c+v_sv_c).
\en
Eq. (\ref{eq:XY}) then follows from Eqs. (\ref{eq:Y2Y1}) and (\ref{eq:Z2Z1}).

\newcommand{\bi}{\bibitem}

\end{document}